\documentclass[conference]{IEEEtran}
\IEEEoverridecommandlockouts
\usepackage[dvipsnames]{xcolor}

\usepackage{cite}
\usepackage{amsmath,amssymb,amsfonts}
\usepackage{algorithmic}
\usepackage{graphicx}
\usepackage{textcomp}
\usepackage{xcolor}
\usepackage{booktabs}
\usepackage{multirow}
\usepackage{threeparttable}
\usepackage{makecell}
\usepackage{todonotes}
\usepackage{colortbl}
\definecolor{coolgrey}{rgb}{0.55, 0.57, 0.67}
\usepackage{tcolorbox}

 \usepackage{url}
\def\BibTeX{{\rm B\kern-.05em{\sc i\kern-.025em b}\kern-.08em
    T\kern-.1667em\lower.7ex\hbox{E}\kern-.125emX}}
\begin{document}

\title{Exploiting Acoustic and Content-Oriented Speaker Verification Attacks Against Multilingual Voice Anonymization}

\author{
\IEEEauthorblockN{
  Ridwan Arefeen\IEEEauthorrefmark{1},
  Ze Li\IEEEauthorrefmark{2},
  Rong Tong\IEEEauthorrefmark{1},
  Ming Li\IEEEauthorrefmark{3},
  Xiaoxiao Miao\IEEEauthorrefmark{4}
}
\IEEEauthorblockA{
  \IEEEauthorrefmark{1}Singapore Institute of Technology, Singapore\\
  \IEEEauthorrefmark{2}Wuhan University, China\\
  \IEEEauthorrefmark{3}The Chinese University of Hong Kong, China\\
  \IEEEauthorrefmark{4}Duke Kunshan University, China
}
\thanks{
2403754@sit.singaporetech.edu.sg, tong.rong@singaporetech.edu.sg, 

xiaoxiao.miao@dukekunshan.edu.cn. 

Corresponding author: Xiaoxiao Miao.
}}
\IEEEaftertitletext{\vspace{-2\baselineskip}}

\maketitle

\begin{abstract}
Attacker ASV systems for voice anonymization have been studied primarily in English, leaving their behavior in multilingual settings largely unexplored. Conventional ASV has shown that both acoustic and contextual information are important for multilingual speaker verification. Inspired by this, we investigate whether the same holds for attacker ASV on anonymized speech. We evaluate both acoustic- and content-oriented attackers on multilingual anonymized speech and construct a multilingual voice-converted dataset to improve cross-lingual generalization. Our results show that attacker effectiveness depends on the linguistic utility of the anonymized speech. Overall, acoustic-oriented attackers achieve better performance. However, when linguistic information is well preserved, the performance gap between content- and acoustic-oriented attackers narrows compared with conditions involving stronger speech distortion. The multilingual voice-converted dataset further improves performance and partially reduces the cross-lingual gap. These findings highlight the need for more comprehensive attacker modeling and evaluation protocols that consider both privacy and utility, rather than relying on a attacker strategy\footnote{Full code and pretrained models and MultiVC Dataset link are available at \texttt{https://github.com/monkeyDarefeen/DAST}.}.
\end{abstract}

\begin{IEEEkeywords}
speaker anonymization, automatic speaker verification, cross-lingual generalization, voice conversion\end{IEEEkeywords}

\section{Introduction}
\label{introduction}
Speech anonymization has emerged as an important research direction in privacy-preserving speech processing, as speech recordings reveal not only linguistic content but also speaker identity and other sensitive attributes.
A central goal is to remove or obfuscate identity-related information without destroying linguistic intelligibility or other utility cues \cite{das2024comparing, VPC2022}.
The privacy protection of an anonymization system is conventionally measured by
how well an attacker's automatic speaker verification (ASV) system can
re-identify the original speaker, typically reported as the equal error rate
(EER). However, existing work \cite{tomashenko2022voiceprivacy,
tomashenko2020introducing} has shown that evaluating against an ASV trained on
real speech is insufficient. A stronger, \emph{semi-informed} attacker instead
trains its ASV model on speech anonymized by the same or similar anonymization system, so that it can
learn to exploit the residual source-speaker information left by that
anonymization. Beyond this, several works further strengthen the attacker, mainly from three distinct perspectives: (i) training on large-scale anonymized or
voice-converted data to trace the source speaker \cite{arefeen2026dastdualstreamvoiceanonymization, mawalim2025fine};
(ii) exploiting residual source-speaker information from other features, such as
linguistic content, duration, and prosody
\cite{franzreb2026content, gaznepoglu2026sayexploitinglinguisticcontent, tomashenko25_interspeech, segreconcat}; and
(iii) improving the ASV architecture, including self-supervised models
\cite{lyu2025fast, 10890291}.

Among these developments, most ASV attacker systems have been developed and assessed largely on
English anonymized speech, with little extension to multilingual settings.  
This is mainly because existing anonymization studies and challenge benchmarks have emphasized English \cite{tomashenko2022voiceprivacy, VPC2022, tomashenko2026third, tomashenko2024first, tomashenko2026privacy}. The latest VoicePrivacy 2026 Challenge (VPC 2026) \cite{miao2026voiceprivacy} introduces
multilingual anonymization (English, German, French, and Spanish), evaluated with a single attacker model shared across all languages. Two main questions, however, remain unexplored. First, the evaluated languages are all Indo-European, so the reported robustness may not reflect how the attacker behaves on structurally distant languages. Second, the challenge results reveal a striking but unexplained pattern: English consistently yields the lowest EER, and the word error rate (WER) of the anonymized speech varies considerably across languages, yet what underlying factor drives this gap remains unclear.

This work thus takes a step further to investigate the bottleneck of the shared attacker across languages. To broaden linguistic diversity beyond the Indo-European family, we deliberately add two structurally distant East Asian languages, Japanese and Chinese, and extend our evaluation to the cross-lingual setting of Track~2 of the VPC~2026 challenge. To probe what actually drives the cross-lingual gap, we compare two
state-of-the-art but fundamentally different attacker paradigms: a WavLM--ECAPA-based
acoustic-oriented attacker \cite{arefeen2026dastdualstreamvoiceanonymization},
the same model used in the
VPC~2026 challenge, and a w2v-BERT-based content-oriented attacker with
multi-scale feature aggregation (MFA) \cite{li2026language}\footnote{We refer to
the former as acoustic-oriented and the latter as content-oriented based on
whether the ASV architecture contains an explicit content-oriented component,
i.e., the BERT component.}.

We first find that both attackers re-identify speakers reliably on the European languages but degrade sharply on the East Asian ones, and that this drop tracks rising automatic speech recognition (ASR) WER, revealing that re-identification degrades in step with ASR utility and suggesting a coupling between linguistic intelligibility and identity disclosure. Comparing the two paradigms under these multilingual conditions, we find the acoustic-oriented attackers are generally more robust than content-oriented ones.

These observations lead to our central hypothesis: the residual identity cues that survive anonymization depend on how well acoustic naturalness and linguistic intelligibility are preserved. If so, exposing an attacker ASV system to diverse, multi-speaker VC data during training should broaden its representations, helping it separate speaker-invariant traits from system-specific anonymization artifacts and narrowing any gap caused by data scarcity rather than by language structure itself. To test this, we construct a large-scale multilingual voice-converted corpus using seven distinct voice-conversion (VC) systems and use it to fine-tune both attackers, yielding domain-adapted variants that partially close the cross-lingual gap.
This shows that the degradation is only partly data-driven, adaptation narrows the gap but a residual portion remains intrinsic to the language itself, which again
highlights the complexity of cross-lingual privacy leakage. These findings highlight the need for more comprehensive attacker modeling that considers both privacy and utility, rather than relying on a single attacker strategy.

\section{Attacker ASV Modeling and Acoustic-Linguistic Vulnerability Analysis} \label{sec:attacker_modeling}
\begin{figure}
\centering
\includegraphics[width=\columnwidth ]{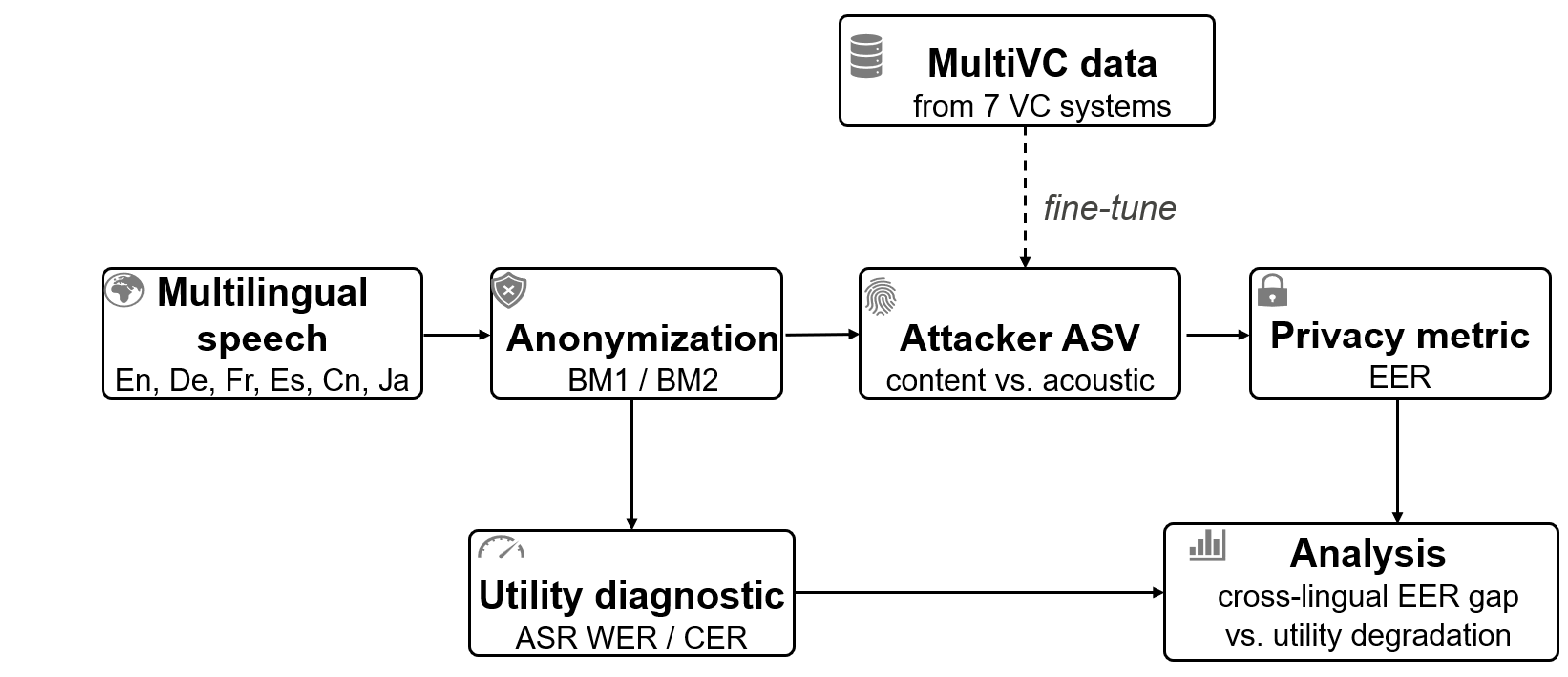}
\caption{Overview of the proposed evaluation.}
\label{fig:overall}
\end{figure}

Figure \ref{fig:overall} gives an overview of our evaluation. Multilingual speech in six languages\footnote{English (En), German (De), French (Fr), Spanish (Es), Chinese (Cn), Japanese (Ja)} is first anonymized by one of two VPC~2026 baselines, BM1 \cite{miao2022language} or BM2 \cite{meyer2024multilingual}. The anonymized speech is then presented to two attacker ASV systems built on fundamentally different paradigms: a WavLM–ECAPA-based acoustic-oriented model \cite{arefeen2026dastdualstreamvoiceanonymization}
 and a w2v-BERT-MFA-based content-oriented model  \cite{li2026language} to quantify re-identification capability, measured by EER.  In parallel, we transcribe the anonymized speech and compute ASR WER/CER as a utility diagnostic, capturing how much intelligibility each pipeline preserves. Our analysis then relates these two axes: the cross-lingual EER gap and utility degradation, to determine whether the drop in re-identification across languages is driven by data scarcity or is intrinsic to the language. To probe the data-scarcity side, we build a large-scale multilingual voice-converted corpus (denoted as MultiVC) generated from seven VC systems) and use it as adaptation data to fine-tune both attackers.

\subsection{Training and Evaluation Data}
\label{train-eval-set}
We perform multilingual attacker system analysis on the official training, development, and test sets of the VPC 2026 Track 2 multilingual anonymization task. For brevity, we report only the test set results in this paper.

To cover language diversity, we construct two auxiliary language datasets. For Japanese, we use Mozilla Common Voice~\cite{ardila2020common} for training (1,545 speakers, 19,015 utterances) and JVS~\cite{takamichi2019jvs} for evaluation (48 speakers, 125,360 trials). For Mandarin Chinese, we use AISHELL-3~\cite{shi21c_interspeech} for both training (340 speakers, 120,418 utterances) and evaluation (30 speakers, 150,648 trials).

\subsection{ Anonymization Baselines}
We take two multilingual anonymization baselines from VPC 2026:

\noindent\textbf{BM1 (HuBERT-based Framework)\cite{miao2022language}:} 
The original speech is decomposed into speaker embeddings, pitch features, and language-independent content representations extracted by a self-supervised pretrained HuBERT model. The speaker embeddings are replaced with those of a target pseudo-speaker, and the pitch and content representations are combined to synthesize the anonymized waveform using the HiFi-GAN neural vocoder.

\noindent\textbf{BM2 (Whisper/Toucan-based Framework)\cite{meyer2024multilingual}:} This architecture extracts speaker embeddings, transcriptions using the pre-trained Whisper model, as well as F0, energy, and phoneme durations. The original speaker embedding is replaced by an artificial embedding generated via a Wasserstein GAN. F0 and energy are randomly perturbed at the phoneme level, and the anonymized speech is synthesized using the IMS Toucan multilingual TTS system\footnote{VPC2026 also includes BM3 in Track~2. Since BM3 follows the same architecture as BM2 and does not incorporate prosody extraction or F0/energy modification, we do not investigate it separately.}.

\subsection{Threat Models}
\label{threat-model}
 We evaluate voice anonymization systems under two attack scenarios. In both cases, the attacker has access to the anonymization system, multiple enrollment utterances for each speaker, and language-matched public speech data, which can be anonymized using the same system to train or adapt the attacker. The enrollment utterances are also anonymized using the same system to match the anonymized trial utterances generated by the user.

\noindent\textbf{Lazy-informed attacker:} The ASV model is not adapted to the corresponding anonymized evaluation data. Instead, it is trained on either original speech or synthetic speech generated by different generative models, as long as these are not the same as those used in the evaluation set.

\noindent\textbf{Semi-informed attacker:} The ASV model is  adapted using anonymized speech generated by the same anonymization system as the evaluation data, resulting in a stronger attacker.

Re-identification capability of the attacker is quantified by the EER, where a lower EER indicates a stronger attacker. Note that VPC2026 focuses primarily on the semi-informed attacker, but in our experiments we consider both attacker settings. The reason is that we aim to investigate a general attacker capable of good generalization across different anonymization approaches, which can be assessed via the lazy-informed EER, while the semi-informed EER further reflects the upper-bound attack strength when the attacker is adapted to the same anonymization system.

\subsection{Attacker Architectures}
\label{attacker-architecture}
 To explore the trade-offs between content-invariant matching and acoustic feature mining under cross-lingual conditions, we evaluate two fundamentally distinct self-supervised models:

\noindent\textbf{Acoustic-oriented (WavLM-ECAPA):} This model features a dual-stream architecture that processes raw waveforms using a frozen WavLM~\cite{9814838} model and spectral features via Mel-filterbanks, with parallel ECAPA-TDNN encoders for mid-level feature fusion. The model is optimized using a three-stage training strategy consisting of speaker pre-training on original speech, diversity-driven domain training on English voice-converted data from the Source Speaker Tracing Challenge (SSTC)~\cite{SSTC} (lazy-informed), and lightweight task-specific fine-tuning on target anonymized data (semi-informed). This is the standard ASV attacker system adopted in VPC 2026. 

\noindent\textbf{Content-oriented (w2v-BERT-MFA):} Built upon a 24-layer Conformer~\cite{gulati2020conformer} architecture, this framework extracts intermediate block-level representations using localized layer adapters and a Multi-scale Feature Aggregation (MFA) module. Pre-trained on a large-scale 4.5-million-hour unlabeled audio corpus via joint contrastive learning and masked language modeling, it is designed to capture text-invariant semantic representations. Parameter-efficient fine-tuning is performed using Low-Rank Adaptation (LoRA)~\cite{hu2022lora} inserted into attention sub-layers. To ensure architectural parity during benchmarking, the pretrained w2v-BERT-MFA model\footnote{\url{https://github.com/ZXHY-82/LI-MSV-TidyVoice2026}} is first fine-tuned with LoRA adapters~\cite{li2025enhancingspeakerverificationw2vbert} on the SSTC~\cite{SSTC} dataset, serving as the lazy-informed attacker. It is then further fine-tuned on the target anonymized data, serving as the semi-informed attacker.

\subsection{Diagnostic Vulnerability Analysis}
\begin{figure}
\centering
\includegraphics[width=\columnwidth ]{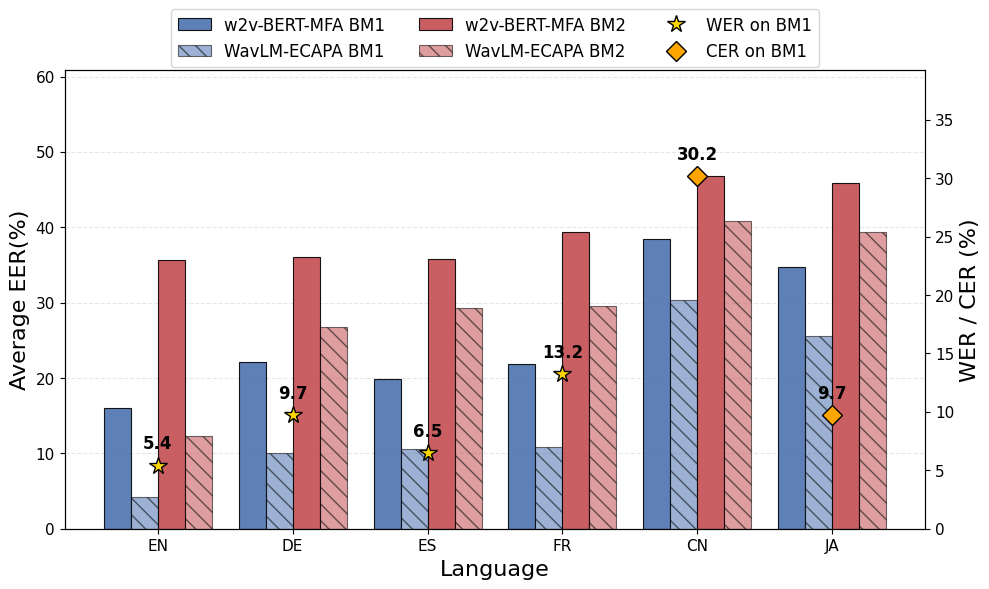}
\caption{Average lazy-informed EER of the content-oriented (w2v-BERT-MFA) and acoustic-oriented (WavLM-ECAPA) attackers across six languages, under the lazy-informed threat model on the BM1 and BM2 baselines. Per-language ASR WER/CER is overlaid (right axis).}

\label{fig:barplot-lazy}
\end{figure}
Figure \ref{fig:barplot-lazy}  presents our first diagnostic view. For each of the six languages, it reports the average lazy-informed EER of the two attackers: content-oriented (w2v-BERT-MFA) and acoustic-oriented (WavLM-ECAPA), against the two anonymization baselines (BM1 and BM2) on multilingual anonymized evaluation sets described in Section \ref{train-eval-set}, with the per-language ASR WER/CER overlaid (with star/diamond). Presenting privacy leakage (EER) and utility loss (WER/CER) on a shared language axis makes it easy to see whether the two move together across languages.  The trends in Fig \ref{fig:barplot-lazy} reveal a cross-lingual generalization bottleneck: while both models reliably recover identities within Indo-European cohorts, their re-identification degrades sharply on the East Asian languages (Chinese and Japanese). 

To examine whether this non-uniform performance is data-driven or tied to linguistic degradation, we conduct a utility analysis, computing WER for the European languages and CER for Chinese and Japanese (where word segmentation is ill-defined) with a Whisper-Large-v3 \cite{radford2023robust}  model transcribing the synthesized audio. The results are reported in Table 1. 

Table \ref{tab:asr_metrics} shows two trends. First, a \emph{baseline synthesis discrepancy}: BM2 yields higher error metrics than BM1 across all languages (e.g., French WER rises from 13.2\% to 44.9\%), indicating that explicit prosodic and F0 extraction imposes more distortion than unified self-supervised embeddings. 
Second, a \emph{typological floor}: this degradation is far larger for logographic and syllabic languages: under BM2, Japanese  reaches a CER of 45.1\% and Chinese a CER of 70.3\%. 

The parallel between utility degradation (Table \ref{tab:asr_metrics}) and re-identification failure (Fig. \ref{fig:barplot-lazy}) points to an acoustic--linguistic coupling. We read this as a coupling rather than a demonstrated causal link: the non-linear transformations of these anonymization systems appear to corrupt speech intelligibility, so when a language's phonetic structure is degraded, the localized identity-bearing cues are diluted, raising the effective difficulty for attackers. 

\begin{table}[t]
\centering
\caption{ASR utility metrics across BM1 and BM2 baselines. CER(\%) is reported for Japanese and Chinese, WER(\%) for the rest.}
\label{tab:asr_metrics}
\begin{tabular}{lcc}
\toprule
Language & BM1 & BM2 \\
\midrule
English (WER)  & 5.4 & 16.0 \\
German (WER)   & 9.7 & 24.2 \\
Spanish (WER)  & 6.5 & 25.6 \\
French (WER)   & 13.2 & 44.9 \\
\midrule
Japanese (CER) & 9.7 & 45.1 \\
Chinese (CER)  & 30.2 & 70.3 \\
\bottomrule
\end{tabular}
\end{table}

\section{Multilingual Voice-Converted Dataset Generation}

\label{sec:dataset_generation}
\begin{figure}
    \centering
    \includegraphics[width=0.68\columnwidth]{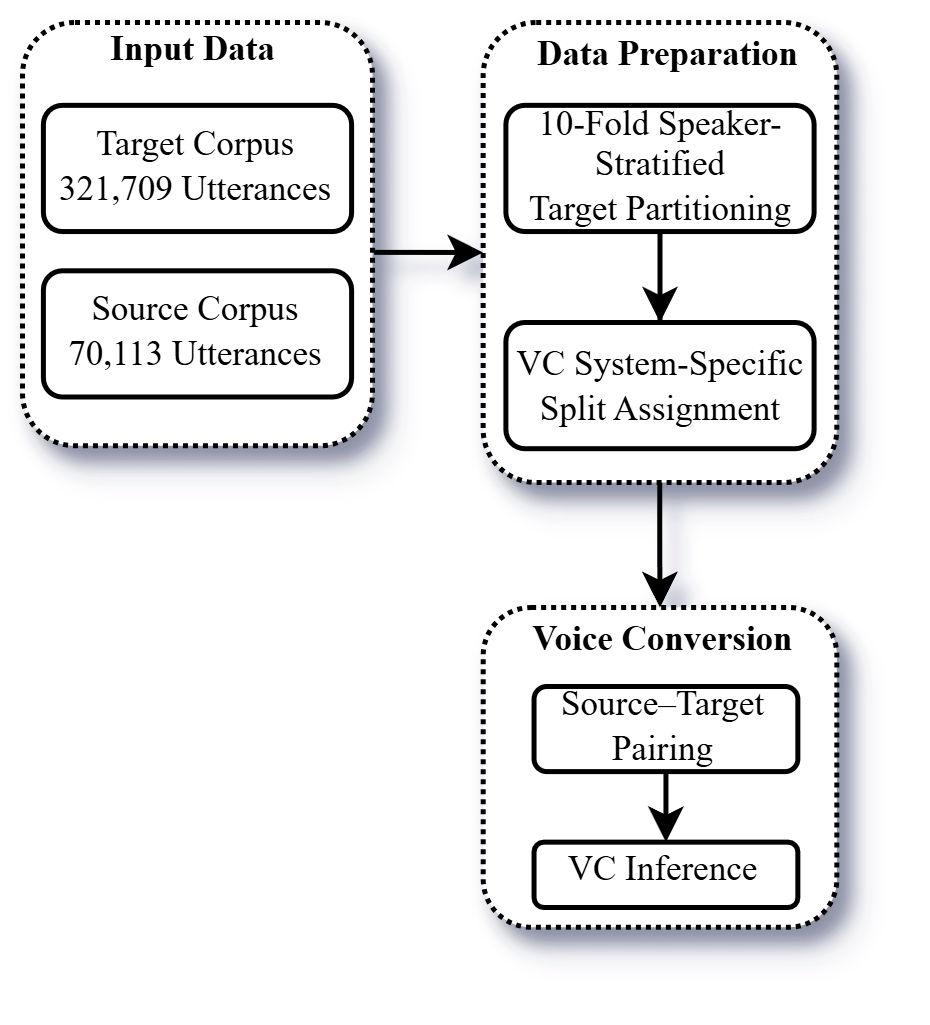}
    \caption{Voice Converted Dataset Pipeline}
    \label{fig:vcpipeline}
\end{figure}
To evaluate whether the cross-lingual performance can be improved by removing data scarcity, we construct a large-scale, multilingual voice-converted corpus. This section details the source and target speech distributions, our speaker-oriented partition mechanics, and the deployment of seven advanced voice conversion architectures used to synthesize our evaluation. Fig. \ref{fig:vcpipeline} shows the overall pipeline for the voice converted dataset generation.


\subsection{Source Corpus}
We construct an expanded multilingual source pool comprising same utterances across all targeted languages (see Fig~\ref{fig:corpus_languages} Right) as from training set mention in section \ref{train-eval-set}. These source utterances serve as the references for our downstream voice conversion pipeline.

\subsection{Target Corpus}

\begin{figure}
    \centering
    \includegraphics[width=1.0\linewidth]{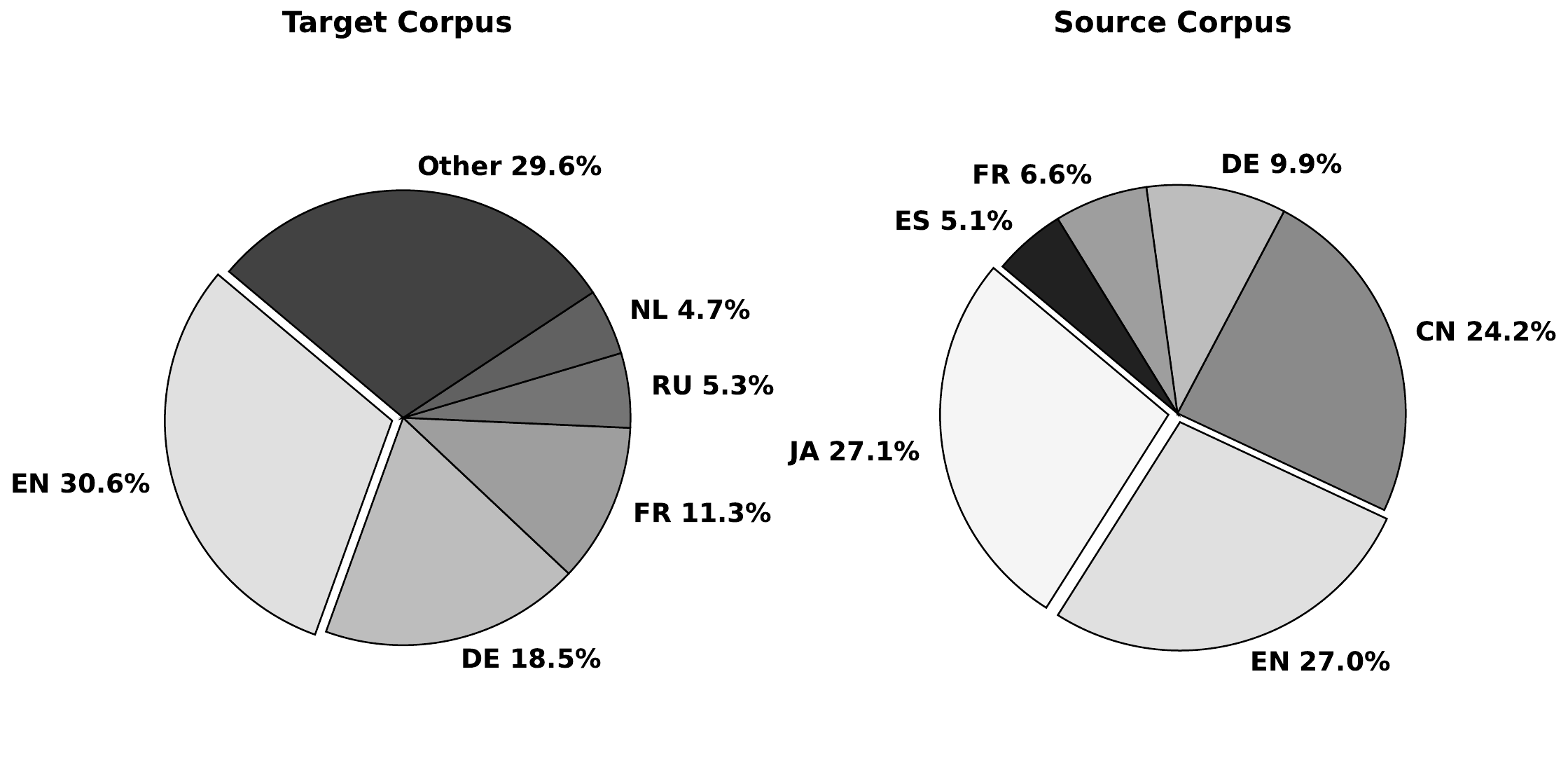}
    \caption{Language distribution of the evaluation corpora. (Left) Target corpus distribution (TidyVoiceX), where the top 5 languages comprise roughly 70\% of the 321,709 utterances. (Right) Source corpus composition, where 70,113 utterances serve as the content reference pool for all seven voice conversion systems.}
    \label{fig:corpus_languages}
\end{figure}

For the target speaker pool, we employ TidyVoiceX from \cite{farhadipour2026tidyvoice}, a diverse multilingual speech corpus containing 321,709 utterances from 4,474 unique speakers across 40 distinct languages. The structural distribution of the top 5 languages (English, German, French, Russian, Dutch) in this target corpus is outlined in (see Fig~\ref{fig:corpus_languages} Left). This corpus provides the necessary acoustic and demographic variance required to create target voice profiles.

\subsection{Split Strategy}
\label{subsec:split_strategy}
We partition the TidyVoiceX corpus into 10 balanced folds using a per-speaker round-robin distribution strategy. For each speaker, utterances are assigned sequentially to splits $1$ through $10$ in cyclic order. When a speaker's total utterance count is not a perfect multiple of 10, the surplus utterances are uniformly distributed via duplication into the remaining splits. This ensures that all 10 folds contain an identical volume of exactly 34,022 utterances while strictly guaranteeing speaker-level stratification; every individual split contains representations from all 4,474 speakers, preventing any speaker profile from appearing exclusively within a single fold.

\subsection{Voice Conversion Systems }

\begin{table}[t]
\centering
\caption{Voice conversion systems and their assigned target splits. Each system converts 34,022 target utterances using 70,113 source utterances, producing 102,066 converted samples per system.}
\label{tab:vc_systems}
\footnotesize
\setlength{\tabcolsep}{6pt}
\begin{tabular}{clc}
\toprule
\textbf{ID} & \textbf{Voice conversion system} & \textbf{Target fold} \\
\midrule
VC-1 & CosyVoice2 \cite{du2024cosyvoice}      & Split 1 \\
VC-2 & Seed-VC \cite{liu2024zeroshotvoiceconversiondiffusion}           & Split 2 \\
VC-3 & YingMusic-SVC \cite{chen2025yingmusic}   & Split 3 \\
VC-4 & FreeVC \cite{10095191}               & Split 4 \\
VC-5 & KDVC \cite{10889083}                  & Split 5 \\
VC-6 & LVC-VC \cite{kang2023end}                & Split 6 \\
VC-7 & TriAAN-VC \cite{10096642}          & Split 7 \\
\midrule
\multicolumn{2}{r}{\textbf{Converted utterances per system}} & \textbf{102,066} \\
\multicolumn{2}{r}{\textbf{Total converted utterances}}       & \textbf{714,462} \\
\bottomrule
\end{tabular}
\end{table}
We utilize seven VC architectures, assigning each to a distinct target split to enforce target-disjoint evaluation and prevent identity cross-contamination. As detailed in Table~\ref{tab:vc_systems}, each architecture processes a fixed allocation of target utterances against the global source pool, yielding a total generated dataset of 714,462 converted samples. Splits 8 through 10 are held in reserve as baseline control sets for future ablation tracking.

\begin{table*}[t]
\centering
\caption{EER (\%) on BM1 and BM2 across six languages. Lower EER indicates a stronger attack. Best results are bolded independently for each block.}

\label{tab:results_combined_transposed}

\setlength{\tabcolsep}{8pt}

\begin{tabular}{l c l l cccccc c}

\toprule
\textbf{Threat Model} & \textbf{ID} & \textbf{ASV Model} & \textbf{Training Data} & \textbf{En} & \textbf{De} & \textbf{Fr} & \textbf{Es} & \textbf{Cn} & \textbf{Ja} & \textbf{Avg.} \\
\midrule

\multirow{4}{*}{\textbf{Lazy-BM1}}
& 1 & w2v-BERT-MFA & +SSTC & 16.76 & 22.20 & 22.73 & 18.84 & 39.76 & 34.86 & \cellcolor{gray!7}25.86 \\
& 2 & w2v-BERT-MFA & +SSTC + MultiVC & 11.108 & 15.336 & 13.141 & 10.565 & 28.160 & 27.670 & \cellcolor{gray!7}17.66 \\
& 3 & WavLM-ECAPA & +SSTC & 5.52 & 11.57 & 11.93 & 8.23 & 31.25 & 26.13 & \cellcolor{gray!7}15.77 \\
& 4 & WavLM-ECAPA & +SSTC + MultiVC & \textbf{2.887} & \textbf{4.526} & \textbf{2.520} & \textbf{1.636} & \textbf{17.727} & \textbf{19.622} & \cellcolor{gray!7}\textbf{8.99} \\

\midrule

\multirow{4}{*}{\textbf{Lazy-BM2}}
& 5 & w2v-BERT-MFA & +SSTC & 35.80 & 34.90 & 39.82 & 34.46 & 46.86 & 45.85 & \cellcolor{gray!7}39.61 \\
& 6 & w2v-BERT-MFA & +SSTC + MultiVC & 35.708 & 34.507 & 38.948 & 33.972 & 46.069 & 44.602 & \cellcolor{gray!7}38.97 \\
& 7 & WavLM-ECAPA & +SSTC & 13.27 & 26.79 & 30.57 & 27.23 & 41.10 & 39.61 & \cellcolor{gray!7}29.76 \\
& 8 & WavLM-ECAPA & +SSTC + MultiVC & \textbf{11.017} & \textbf{21.053} & \textbf{24.781} & \textbf{18.443} & \textbf{37.884} & \textbf{35.814} & \cellcolor{gray!7}\textbf{24.83} \\

\midrule
\midrule

\multirow{4}{*}{\textbf{Semi-BM1}}
& 9 & w2v-BERT-MFA & +SSTC + BM1 & 4.142 & 5.684 & 6.086 & 4.313 & 14.364 & 15.206 & \cellcolor{gray!7}8.30 \\
& 10 & w2v-BERT-MFA & +SSTC + MultiVC + BM1 & 3.991 & 5.409 & 5.463 & 4.210 & 14.592 & 14.867 & \cellcolor{gray!7}8.09 \\
& 11 & WavLM-ECAPA & +SSTC + BM1 & 2.887 & {1.081} & 1.091 & 1.217 & 7.875 & 11.496 & \cellcolor{gray!7}4.27 \\
& 12 & WavLM-ECAPA & +SSTC + MultiVC + BM1 & \textbf{2.786} & \textbf{0.914} & \textbf{1.071} & \textbf{0.992} & \textbf{6.285} & \textbf{9.910} & \cellcolor{gray!7}\textbf{3.66} \\

\midrule

\multirow{4}{*}{\textbf{Semi-BM2}}
& 13 & w2v-BERT-MFA & +SSTC + BM2 & 48.131 & 32.206 & 38.121 & 31.482 & 46.279 & 44.034 & \cellcolor{gray!7}40.04 \\
& 14 & w2v-BERT-MFA & +SSTC + MultiVC + BM2 & 26.270 & 30.604 & 35.433 & 29.207 & 45.141 & 42.633 & \cellcolor{gray!7}34.88 \\
& 15 & WavLM-ECAPA & +SSTC + BM2 & 7.185 & 10.561 & 13.389 & 7.139 & 28.992 & 28.994 & \cellcolor{gray!7}16.04 \\
& 16 & WavLM-ECAPA & +SSTC + MultiVC + BM2 & \textbf{7.013} & \textbf{9.899} & \textbf{12.610} & \textbf{6.650} & \textbf{26.786} & \textbf{28.409} & \cellcolor{gray!7}\textbf{15.23} \\
\bottomrule
\end{tabular}
\end{table*}

\subsection{Pair Generation}
 
To synthesize pairs for each VC system, we map each of the 34,022 target style references to three distinct source content utterances, yielding $34,\!022 \times 3 = 102,\!066$ conversion utterances per system. The pairing framework follows a two-phase allocation strategy engineered to maximize global source coverage:
\begin{enumerate}
    \item \textbf{Guaranteed Coverage (Phase~1):} The entire pool of 70,113 source utterances (Table~\ref{tab:vc_systems}) is randomized and assigned to the initial 70,113 pair slots, ensuring every source language token acts as a content reference at least once.
    \item \textbf{Stochastic Backfill (Phase~2):} The remaining $102,\!066 - 70,\!113 = 31,\!953$ slots are populated via uniform random sampling from the source pool with replacement.
\end{enumerate}

This strategy ensures that the entire source corpus is represented in every conversion set. Across systems, the average source usage ranges from $1.35\times$ to $1.46\times$, with a maximum cap of 6 uses per source utterance. The distribution of source usage follows a long-tail pattern: approximately 70\% of sources are utilized once, 20\% twice, and the remainder three or more times.

\section{Experiment}

We first evaluate lazy-informed attackers on multilingual test sets. Since the training data combine multiple languages, we expect improved cross-lingual generalization ability against different anonymization systems. This setting allows us to study the robustness of both acoustic-oriented (WavLM-ECAPA) and content-oriented (w2v-BERT-MFA) ASV models under different synthetic training conditions, including monolingual (SSTC) and multilingual (MultiVC) data.

We then evaluate semi-informed attackers, where the lazy-informed models are further fine-tuned on the target anonymization baselines, i.e., BM1 or BM2. We further analyze the bottlenecks of the multilingual attacker system.

\subsection{Training Setup for the Attacker Models}

As introduced in Section~\ref{threat-model} and Section~\ref{attacker-architecture}, we consider two attacker scenarios, i.e., lazy-informed and semi-informed, under two ASV architectures, i.e., WavLM-ECAPA and w2v-BERT-MFA.

During lazy-informed attacker training, WavLM-ECAPA follows the training strategy in \cite{arefeen2026dastdualstreamvoiceanonymization}. The content-oriented w2v-BERT-MFA framework is fine-tuned using LoRA, following the architectural constraints in \cite{li2025enhancingspeakerverificationw2vbert}. Both architectures are trained on SSTC (English voice conversion data) or MultiVC, where training data from multiple languages are combined to improve generalization.

During semi-informed attacker training, we follow the official VPC 2026 ASV fine-tuning script to optimize both ASV architectures under target anonymization data, i.e., BM1 and BM2. The models are further fine-tuned on the corresponding target dataset for each scenario.

\subsection{Experimental Results}
\subsubsection{Results on Lazy-informed Attackers}
We first analyze the lazy-informed attacker results, where the top portion of Table~\ref{tab:results_combined_transposed} reports the EER on multilingual test sets under both BM1 and BM2 anonymization settings.

From the training data perspective, incorporating multilingual data consistently improves performance. For BM1 (ID1–ID4), the average EER is reduced from 25.86\% to 17.66\% for w2v-BERT-MFA, and from 15.77\% to 8.99\% for WavLM-ECAPA. Similar trends are observed on BM2 (ID5–ID8), where the average EER decreases from 39.61\% to 38.97\% for w2v-BERT-MFA, and from 29.76\% to 24.83\% for WavLM-ECAPA. These results demonstrate that multilingual training data provides consistent but moderate gains across both architectures.

However, the improvement brought by multilingual training is substantially smaller on BM2 compared to BM1. This suggests that stronger anonymization significantly limits the benefit of cross-lingual generalization. A possible explanation is that BM2 employs ASR–TTS-based speech generation, which introduces stronger prosodic distortion and partial semantic degradation (see WER/CER results on BM2 in table \ref{tab:asr_metrics}). As a result, voice conversion-based training data (e.g., SSTC, MultiVC) becomes less effective in preserving or recovering speaker-discriminative cues. This hypothesis is further investigated in the semi-informed evaluation.

Comparing ASV architectures, WavLM-ECAPA consistently outperforms w2v-BERT-MFA across both BM1 (ID1–2 vs. ID3–4) and BM2 (ID5–6 vs. ID7–8). This indicates that acoustic-oriented representations are more robust under anonymized conditions, whereas content-oriented models such as w2v-BERT-MFA rely more heavily on semantic consistency. When anonymization distorts semantic information, the effectiveness of content-based speaker modeling is reduced, making acoustic representations more reliable for attacker construction.

\subsubsection{Results on Semi-informed Attackers}

We further evaluate semi-informed attackers by fine-tuning each lazy-informed model on anonymized speech generated by the corresponding target anonymization system. The results are reported in the lower part of Table~\ref{tab:results_combined_transposed}.

Unlike in the lazy-informed setting, the relative effectiveness of different attackers depends on the utility of the anonymized speech. On BM1, WavLM-ECAPA with MultiVC + SSTC + BM1 (ID12) achieves the best performance, yielding the lowest average EER of 3.66\%. WavLM-ECAPA also outperforms w2v-BERT-MFA on BM2 (ID15--16 vs.\ ID13--14), with the best result obtained using MultiVC augmentation (ID16, 15.23\% average EER). However, the performance gap between the two attacker types differs substantially across BM1 and BM2.

This difference is consistent with the utility results in Table~\ref{tab:asr_metrics}. BM1 achieves lower WER/CER across all languages, indicating better preservation of linguistic information. Under this condition, the content-oriented attacker performs more closely to the acoustic-oriented attacker. In contrast, BM2 introduces stronger speech distortion, which degrades the linguistic information available to the content-oriented attacker and widens the performance gap in favor of WavLM-ECAPA. MultiVC augmentation further improves attack performance in most settings.

Overall, acoustic-oriented attackers achieve better performance across both anonymization settings. However, the relative effectiveness of content-oriented attacks depends more strongly on speech utility: when linguistic information is well preserved, their performance approaches that of acoustic-oriented attacks, whereas stronger distortion increases the advantage of acoustic-oriented attacks.

\section{Conclusion}

This study investigated the bottleneck of cross-lingual generalization for attacker ASV systems against voice anonymization. To determine whether the performance gap across languages is driven by data scarcity or linguistic intelligibility, we compared an acoustic-oriented dual-stream model (WavLM-ECAPA) with a content-oriented unified semantic model (w2v-BERT-MFA). Our experiments establish a clear coupling between speech utility (measured by WER/CER) and re-identification vulnerability; as well in semi-informed scenario, content-oriented models (w2v-BERT-MFA) gives competitive results when semantic structure is preserved, but under heavily distorted, high-WER conditions, acoustic-oriented architectures (WavLM-ECAPA) demonstrate far superior robustness. This indicate that the optimal attackers outcome depends on how well the target anonymization system preserves semantic information. Furthermore, while fine-tuning on our constructed multi-speaker voice-converted corpus (MultiVC) mitigates data scarcity and partially narrows the cross-lingual EER gap, a residual, intelligibility-bound gap inherently persists. Consequently, these observations highlight the need for multifaceted attacker modeling and evaluation protocol that can weigh privacy preservation againts utility degradation.

\section{Acknowledgments}
This work is supported by the Singapore Ministry of Education (MOE) Academic Research Fund (AcRF) Tier 1 grant R-R13-A405-0005.

\section{Generative AI Use Disclosure}
Large language models (LLMs) were used in a limited capacity as auxiliary tools to assist with the language polishing of the manuscript. All research ideas, methodological design, experiments, analyses, and final writing decisions were conducted and verified by the authors.

\bibliographystyle{IEEEtran}
\bibliography{mybib}

\end{document}